# Assessing Google Correlate Queries for Influenza H1N1 Surveillance in Asian Developing Countries

Xichuan Zhou[1*], Qin Li[2], Han Zhao[2], Shengli Li [1], Lei Yu [1], Fang Tang [1], Shengdong Hu [1], Guojun Li[4], Yujie Feng[3]

1   College of Communication Engineering, Chongqing University, Chongqing, China

2   Chongqing Centers for Disease Control and Prevention, Chongqing, China

3   The Third Military Medical University, Chongqing, China

4   Chongqing Communication Institute, Chongqing, China

E-mail: zhouxichuan@cqu.edu.cn

## Abstract

So far, Google Trend data have been used for influenza surveillance in many European and American countries; however, there are few attempts to apply the low-cost surveillance method in Asian developing countries. To investigate the correlation between the search trends and the influenza activity in Asia, we examined the Google query data of four Asian developing countries. The search data of 26 queries from China, India, Malaysia and Philippines were examined. Our study was based on a publicly available Google service known as Google Correlate. National H1N1 influenza virological data were obtained from the World Health Organization. For each country, Google Correlate calculated and ranked the Pearson correlation coefficients between the influenza data and the search trends of all the queries in Google's database. We examined the weekly updated trend data from January 4, 2004 to December 29, 2013. The highest correlations between the H1N1 influenza virological data and the Google Correlate query data for respective countries ranged from 0.81 to 0.96 ($p < 0.05$). The influenza-related queries which achieved the highest correlation for each country were *Lian-Hua-Qing-Wen capsule* for China ( a traditional influenza medicine, $r = 0.86, p < 0.05$), *H1N1 vaccine* for India ($r = 0.81, p < 0.05$), *H1N1 Malaysia* for Malaysia ($r = 0.88, p < 0.05$) and *A H1N1* for Philippines ($r = 0.96, p < 0.05$). This study indicated that the search queries provided by Google Correlate could be used as a complementary source of data for H1N1 influenza surveillance in the examined Asian developing countries.

## Introduction

Asia had a series of influenza epidemics in the last a few years, causing serious concern around the world. Effective surveillance approach was the first defense line against emerging epidemics; however, traditional surveillance networks relied on case reporting, which involved days of delay in case diagnosis and confirmation. Moreover, some developing countries in Asia lacked the resources to build traditional surveillance networks. Before the 2009 H1N1 influenza pandemic, researchers showed that the Internet searches of influenza-related queries were good indicators of influenza activity [1–4]. Although the concept of affordable real-time surveillance was promising, few studies succeeded to extend the approach to Asian developing countries due to insufficient search data [5–10]. Luckily, the situation was changing fast. The Internet access rate in developing countries had doubled since 2009 [11].

In November 2008, Google started Google Flu service, which used a computational search query model to estimate influenza activity. So far, the Google Flu has covered most European and American countries [12–22], however, there is currently no Google Flu in Asian developing countries. Recently, academic researchers became interested in extending the Google Flu method to Asian countries and regions. Yuan examined the query data of Baidu search engine for the purpose of influenza surveillance in China [10]. Kang focussed on regional-level surveillance, and found that the influenza virological surveillance data of Guangdong province in China was correlated with provisional Google search trend of the Chinese term of influenza A ($r = 0.64, p < 0.05$) [6]. Cho examined the Google trends in South Korea and found that the Google was correlated with the influenza-like illness trend ($r = 0.53, p < 0.05$) [5]. These inspiring work motivated us to examine if the Google search trend could be used for monitoring national-level influenza activity in Asian developing countries.

One challenge of search-based influenza surveillance is query selection. Due to restricted accessability of search data, academic researchers generally selected the influenza-related queries by domain knowledge. Cho conducted a public survey of 100 patients and collected 12 influenza related queries for surveillance [5]. Kang selected 7 influenza related queries including the names and symptoms of influenza by common knowledge [6]. Recently, the Google Correlate service was released [23], which was a query correlation analysis tool designed to provide the top ranking queries correlated with given data series. In this research, we analyzed the H1N1 influenza related queries provided by Google Correlate for the purpose of influenza surveillance in Asian developing countries.

## Methods

We examined four Asian countries, i.e. China, India, Malaysia and Philippines based on available surveillance data. We analyzed the official virological data of the H1N1 influenza published by the World Health Organization (WHO) [24]. The virological data spanned a period from January 2004 to December 2013 and were of a weekly resolution (Figure 1). The data were provided to the WHO remotely by National Influenza Centers of the Global Influenza Surveillance and Response System (GISRS) and other national influenza reference laboratories collaborating with GISRS.

We used the Google Correlate tool to assess the Pearson correlation between the H1N1 influenza surveillance data and the search trends of different queries. Google Correlate was a publicly available tool enabling people to find the queries with the most similar search pattern to an uploaded data series, which in our case was the national H1N1 influenza virological data. For each selected country, Google Correlate computed the Pearson correlation coefficient between the virological data and the search-volume of every query submitted from the examined country. Then, Google Correlate ranked the queries with respect to the correlation coefficients. For each examined country,

Google Correlate listed the top correlated queries. The search data of the top-ranking queries could be downloaded and Google Correlate normalized the data with zero of 1.

We analyzed the 10-year-overall and yearly correlation coefficients of 21 influenza-related queries provided by Google Correlate. We also examined if the Pearson correlation could be improved by shifting the search query data forward and afterward. For this lag-correlation analysis, we used the corrcoef function of the Matlab software to calculate the Pearson correlation coefficients. Significance was set at $p < 0.05$ for all the experiments and strong correlation was defined as a correlation coefficient $r \geq 0.70$.

## Results

Table 1 listed the Google Correlated queries which were strongly correlated with the national H1N1 influenza surveillance data. The 21 queries in table 1, composed of words and blank spaces, were in different local languages for respective countries. Google treated different combinations of the same words as different queries. For the four examined Asian countries, we found that, among all the queries submitted to Google, the top queries correlated with the H1N1 influenza virological data included the names, symptoms and protective means of the H1N1 influenza disease. For all the 21 examined queries, the overall correlation for the period from January 2004 to December 2013 were relatively high, ranging from 0.70 to 0.95 ($p < 0.05$). For India, among all the queries submitted in India, the query with the highest correlation was H1N1 vaccine ($r = 0.81, p < 0.05$). The query terms of fluvir, swine flu treatment, swine flu vaccine and vaccine side effect were also found strongly correlated with Indian H1N1 influenza virological data ($r = 0.73, 0.72, 0.70, 0.74, p < 0.05$). For China, the name of a traditional Chinese medicine for influenza was found the sixth highest ranking query with respect to Pearson correlation coefficient. The name of the traditional medicine had two spellings, both of which were strongly correlated with Chinese H1N1 influenza virological data ($r = 0.86, 0.85, p < 0.05$). The Chinese word for influenza A was also found strongly correlated with Chinese H1N1 influenza infections ($r = 0.79, p < 0.05$). For Philippines, the Google search trend for AH1N1 was found the second highest ranking query with correlation coefficient $r = 0.90$ ($p < 0.05$). Four other queries including H1N1, A H1N1, influenza A and AH1N1 virus were found strongly correlated with Philippine H1N1 influenza virological data ($r = 0.80, 0.85, 0.71, 0.79, p < 0.05$). For Malaysia, the search trend for the term H1N1 Malaysia was found the third highest ranking query with correlation coefficient $r = 0.88$ ($p < 0.05$). Eight other queries including H1N1, H1N1 virus, influenza A, sanitizer and H1N1 symptoms were found strongly correlated with Malaysian H1N1 influenza virological data ($r = 0.85, 0.80, 0.77, 0.72, 0.80, p < 0.05$).

As shown in table 1, we assessed whether the H1N1 influenza virological data had a higher correlation with the Google trend data proceeding or lagging a period of time. Both the search trend data and the virological data were updated on a weekly basis. Form January 2004 to December 2013, 16 out of 21 influenza-related queries had the highest correlation when the Google search data had zero week of proceeding time. The rest 10 queries had higher correlation coefficients with search volume data proceeding one or two weeks of time. Different countries had different results in terms of optimal proceeding time. Specifically, two countries had the highest average correlation coefficients with zeroweek proceeding time, including India (r = 0.74, p < 0.05) and Malaysia (r = 0.80, p < 0.05). China and Philippines had the highest average correlation coefficients of 0.83 (p < 0.05) and 0.95 (p < 0.05) with one-week and two-week proceeding time respectively.

Our analysis used 520 weeks of data from 2004 to 2013. In table 2, we assessed the yearly correlation between the examined queries and the H1N1 influenza surveillance data with zero-week lag. Since there were very few H1N1 influenza cases before 2009, the Pearson correlation calculation was not applicable for the years from 2004 to 2008. Therefore, we only listed the results from 2009 to 2013 in table 2. For the examined five countries, the overall correlation for the whole period of ten years were relatively high for all the examined queries, ranging from 0.70 to 0.95 (p < 0.05) with zero-week lag. In terms of yearly correlation, most queries (23 out of 26) had higher yearly correlation coefficients in 2009, ranging from 0.70 to 0.96 (p < 0.05). Figure 2 showed the average yearly correlation of different queries for each examined country. As one can see, though the correlation for most analyzed queries were positive, the correlation coefficients were relatively lower in non-pandemic years.

## Discussion

Though people's search behavior varied significantly from country to country due to different local languages and levels of education, we found that the influenza related queries provided by Google Correlated were strongly correlated with national H1N1 influenza virological data for the examined Asian developing countries. Prior studies of using Google search data for influenza surveillance in Asian countries and regions only reported moderate correlation coefficients between Google query data and the influenza surveillance data of South Korea of and Southern China (r < 0.65, p < 0.05) [5,6]. Though the data sources were different, our paper reported generally higher overall correlation between Google query data and the national influenza surveillance data, ranging from 0.70 to 0.95 (p < 0.05). One possible reason may be because Cho and Kang chose the influenza-related queries by domain knowledge, which could only be a proportion of influenza-related queries in the Google database, resulting in possible omissions of good indicators of influenza activity, e.g. the query of Lian-Hua-Qing-Wen capsule in China (r = 0.85, p < 0.05).

The correlation analysis in our research showed that, for China, Malaysia and Philippine, the correlation of some queries can be improved with the search time-series proceeding the surveillance data for one or two weeks (Table 1). This phenomenon was consistent with some studies [2, 5, 25], which suggested that the influenza surveillance system based on these queries might detect the change of influenza activity before traditional virological surveillance networks.

There were several limitations to this study. First, the correlation coefficients of non-pandemic years were relatively lower compared with H1N1 influenza pandemic period in 2009 (Figure 2). The main reason for the lower correlation was because the queries we examined were selected by ranking the overall correlation coefficients of ten-year data, which were significantly affected by the huge spike in influenza activity in 2009. The second limitation was the possible error in influenza virological data. Study showed that the official numbers for the H1N1 influenza pandemic in 2009 were likely underestimated [26]. The third limitation of our research was the lack of search data, which was the main reason why this research didn't cover more Asian countries such as South Korea. So far, Google Correlate only provided the query data of 53 countries.

In conclusion, we found that the Google search data of influenza related queries were correlated with national H1N1 influenza virological data in the examined Asian developing countries. By taking advantage of readily available data essentially submitted by millions of individuals, Google search engine could provide useful and low-cost influenza surveillance for the developing world. The results of our research suggested that the the search trend of queries provided by Google Correlate can be used as complementary data for influenza surveillance. However, the relatively lower correlation for non-pandemic period could introduce error in predictive models like the Google Flu. More research is required to design more reliable and accurate models.

# Figure Legends

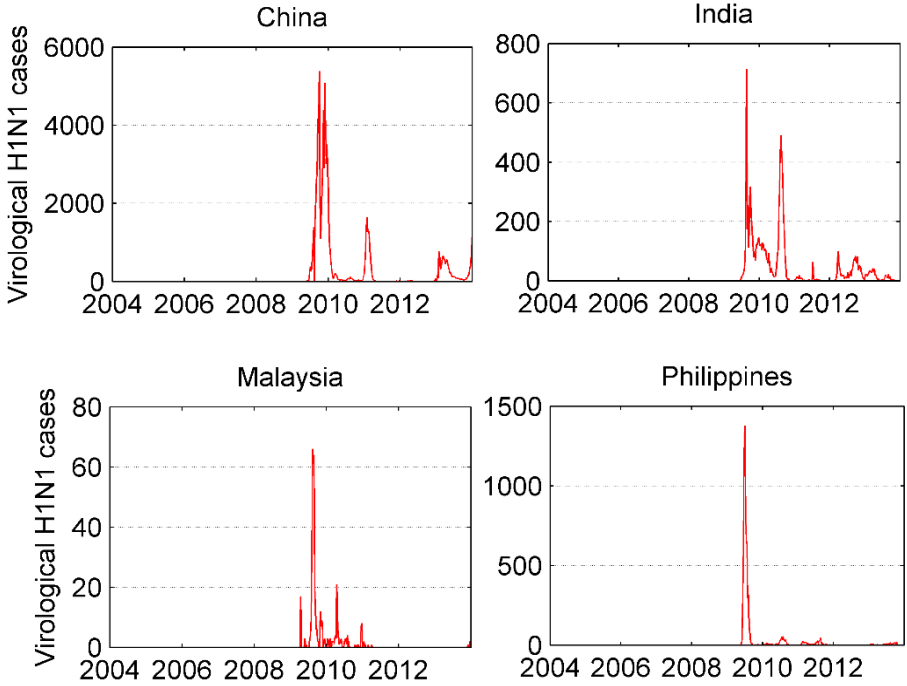

**Figure 1. Time series plots of the H1N1 influenza virological data for the examined Asian countries, from January 2004 to December 2013.**

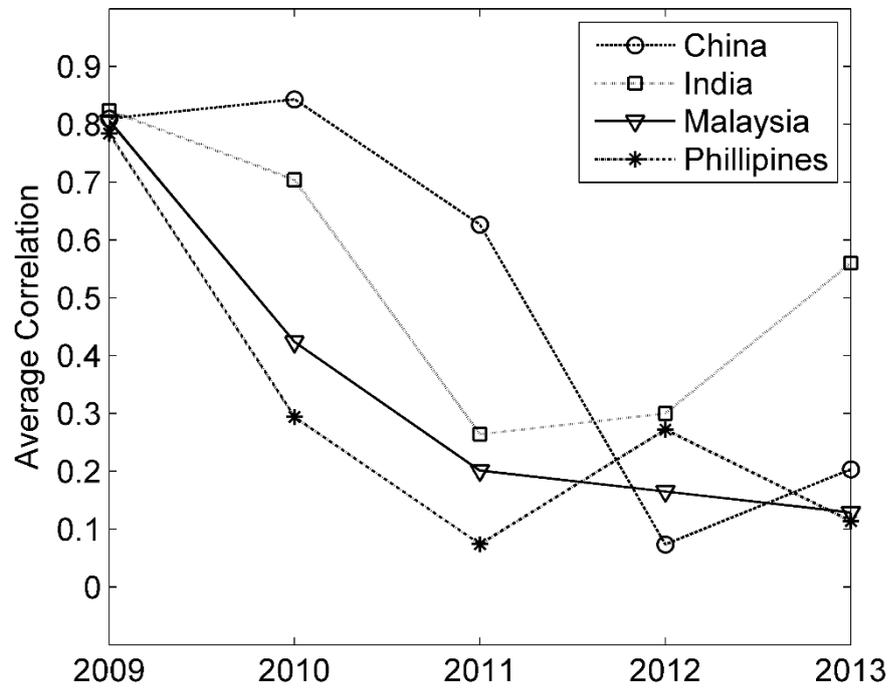

**Figure 2. The average yearly correlation coefficients for each examined country.**

# Tables

**Table 1. Google Correlate queries which were strongly correlated with the H1N1 virological trend for the period from January 2004 to December 2013.**

| Country | Queries | lagging weeks | | proceeding weeks | |
|---|---|---|---|---|---|
| | | 0 week | 1 week | 1 weeks | 2 week |
| China | 连花清瘟胶囊 (Lian-Hua-Qing-Wen capsule) | 0.86* | 0.81 | 0.85 | 0.81 |
| | 莲花清瘟胶囊 (Lianhua-Qing-Wen capsule) | 0.85* | 0.79 | 0.84 | 0.81 |
| | 甲流感 (influenza A) | 0.79 | 0.73 | 0.82* | 0.74 |
| India | H1N1 vaccine | 0.81* | 0.78 | 0.69 | 0.61 |
| | fluvir | 0.73* | 0.63 | 0.70 | 0.56 |
| | swine flu treatment | 0.72* | 0.53 | 0.65 | 0.43 |
| | swine flu vaccine | 0.70* | 0.54 | 0.64 | 0.32 |
| | vaccine side effects | 0.74* | 0.54 | 0.73 | 0.41 |
| Malaysia | penyakit H1N1 (H1N1 disease) | 0.81 | 0.80 | 0.84* | 0.72 |
| | H1N1 Malaysia | 0.88* | 0.69 | 0.76 | 0.86 |
| | H1N1 | 0.85* | 0.75 | 0.79 | 0.60 |
| | H1N1 symptoms | 0.80 | 0.77 | 0.80* | 0.68 |
| | sanitizer | 0.72 | 0.75 | 0.79* | 0.71 |
| | H1NI | 0.84* | 0.73 | 0.78 | 0.67 |
| | H1N1 virus | 0.80* | 0.70 | 0.79 | 0.78 |
| | influenza A | 0.77* | 0.69 | 0.72 | 0.53 |
| Philippines | AH1N1 | 0.90 | 0.89 | 0.96* | 0.96 |
| | A H1N1 | 0.85 | 0.72 | 0.96* | 0.95 |
| | influenza A | 0.71 | 0.64 | 0.89 | 0.96* |
| | AH1N1 virus | 0.79 | 0.62 | 0.93 | 0.96* |
| | H1N1 | 0.80 | 0.72 | 0.92 | 0.95* |

We selected the above Google Correlate queries by two criteria: 1) the search trend data with zero-week lag were strongly correlated with the influenza surveillance data ($r \geq 0.70, p < 0.05$), 2) the queries were practically related with the H1N1 influenza disease.

**Table 2. Yearly correlation analysis between H1N1 virological surveillance data and the Google query data from January 2009 to December 2013.**

| Country | Queries | 2009 | 2010 | 2011 | 2012 | 2013 |
|---|---|---|---|---|---|---|
| China | 连花清瘟胶囊 (Lian-Hua-Qing-Wen capsule) | 0.85* | 0.82 | 0.36 | 0.06 | 0.24 |
| | 莲花清瘟胶囊 (Lianhua-Qing-Wen capsule) | 0.82* | 0.77 | 0.82 | 0.15 | 0.12 |
| | 甲流感 (influenza A) | 0.76 | 0.94* | 0.70 | 0.01 | 0.25 |
| India | H1N1 vaccine | 0.89* | 0.65 | 0.30 | 0.16 | 0.63 |
| | fluvir | 0.66 | 0.78* | 0.05 | 0.64 | 0.74 |
| | swine flu treatment | 0.91* | 0.77 | 0.41 | 0.29 | 0.56 |
| | swine flu vaccine | 0.84* | 0.70 | 0.29 | 0.29 | 0.61 |
| | vaccine side effects | 0.82* | 0.62 | 0.27 | 0.12 | 0.26 |
| Malaysia | penyakit H1N1 | 0.80* | 0.38 | 0.06 | NA | 0.13 |
| | H1N1 Malaysia | 0.87* | 0.59 | 0.09 | NA | 0.07 |
| | H1N1 | 0.84* | 0.46 | 0.26 | NA | 0.17 |
| | H1N1 symptoms | 0.78* | 0.42 | 0.31 | NA | 0.13 |
| | sanitizer | 0.80* | 0.20 | 0.38 | NA | 0.37 |
| | H1NI | 0.83* | 0.51 | 0.11 | NA | 0.10 |
| | H1N1 virus | 0.78* | 0.48 | 0.05 | NA | 0.04 |
| | influenza A | 0.75* | 0.35 | 0.35 | NA | 0.02 |
| Philippines | AH1N1 | 0.88* | 0.53 | -0.02 | 0.20 | 0.30 |
| | A H1N1 | 0.82* | 0.31 | 0.06 | 0.39 | 0.12 |
| | influenza A | 0.70* | 0.03 | 0.22 | 0.18 | 0.06 |
| | AH1N1 virus | 0.76* | 0.29 | -0.01 | 0.18 | 0.02 |
| | H1N1 | 0.76* | 0.31 | 0.06 | 0.41 | 0.07 |

NA: Not applicable.

Since there were very few H1N1 influenza cases before 2009, the Pearson correlation calculation was not applicable before 2009. Malaysia reported no H1N1 influenza cases in 2012, therefore the 2012 yearly correlation computation for Malaysia were not applicable.